\documentclass[10pt]{iopart}

\usepackage[latin1]{inputenc}
\usepackage[dvips]{epsfig}
\usepackage{amsfonts}

\newcommand{\eps}{\varepsilon}
\newcommand{\vphi}{\varphi}
\newcommand{\la}{\lambda}

\begin{document}

\title[Deformation of a Cylinder under Tidal Gravitational Forces]
{Analytical Solution for the Deformation of a Cylinder under Tidal Gravitational Forces}

\author{S Scheithauer and C L\"ammerzahl}

\address{ZARM, University of Bremen
Am Fallturm, 28359 Bremen, Germany}
\ead{scheithauer@zarm.uni-bremen.de}

\begin{abstract}
Quite a few future high precision space missions for testing Special and General Relativity 
will use optical resonators which are used for laser frequency stabilization. These devices 
are used for carrying out tests of the isotropy of light (Michelson-Morley experiment) and of 
the universality of the gravitational redshift. As the resonator frequency not only depends on 
the speed of light but also on the resonator length, the quality of these measurements is very 
sensitive to elastic deformations of the optical resonator itself. As a consequence, a detailed 
knowledge about the deformations of the cavity is necessary. Therefore in this article we investigate the 
modeling of optical resonators in a space environment. Usually for simulation issues 
the Finite Element Method (FEM) is applied in order to investigate the influence of disturbances 
on the resonator measurements. However, for a careful control of the numerical quality of FEM 
simulations a comparison with an analytical solution of a simplified resonator model is beneficial. 
In this article we present an analytical solution for the problem of an elastic, isotropic, homogeneous 
free-flying cylinder in space under the influence of a tidal gravitational force. The solution is 
gained by solving the linear equations of elasticity for special boundary conditions. The applicability 
of using FEM codes for these simulations shall be verified through the comparison of the analytical 
solution with the results gained within the FEM code.

\end{abstract}

\maketitle


\section{Motivation}

Special (SR) and General Relativity (GR) are two of the most important theories and theoretical frames of modern physics. They are the basis for the understanding of space and time and thus for the underlying physical structure of any other theory. The interest in testing the fundamentals of SR and GR has grown enormously over the last years as all presently discussed approaches to 
quantum gravity predict tiny violations of SR and GR. 

The technological improvements of the last decades have provided scientists with high precision measurement equipment such as optical resonators. In optical resonators laser locking is used to define stable optical frequencies. The resonance frequency of the locked lasers is given by $\nu = m c/L$ where $c$ is the speed of light, $L$ the resonator length $L$, and $m$ the mode number. 

Optical resonators have been used recently to test one of the pillars of Special Relativity, namely the isotropy of the speed of light \cite{Muelleretal03, Wolfetal03} as well as of the universality of the gravitational redshift \cite{Braxmaieretal02}. In doing so, two laser beams are locked to two orthogonally oriented resonators. An anisotropic speed of light would lead to a beat of the frequencies during a rotation of this setup. Due to the importance of this type of experiments one looks for ways to improve that. One option for this is to carry out these experiments in space, as planned in the OPTIS mission \cite{laemmerzahl04:_optis} or with SUMO on the ISS \cite{Laemmerzahletal04}. 

Although many of the disturbances acting on a resonator can be minimized by means of an  appropriate satellite control system, some intrinsic disturbances cannot be eliminated as a matter of principle and distort the resonator shape leading to a systematic frequency shift. 
In particular the tidal gravitational force \footnotemark \ which acts through every extended body cannot be eliminated 
by choosing an appropriate frame and, thus, will induce distortions on the resonator. 
\footnotetext{In space and engineering sciences the tidal gravitational force is often 
referred to as 'gravity gradient'.}

We give a rough estimate of the expected effect of the tidal gravitational force on a freely moving cube of length $L$. If the position of the cube is at a distance $R$ from the center of the Earth, then the difference of the Earth's acceleration on the top and bottom of the cube is $\Delta a = (\partial^2 U/\partial r^2) L$, where $U$ is the Earth's Newtonian potential $U = GM_\oplus/R$. For an orbit with $R = 10000\;{\rm km}$ and a typical resonator length of $L = 5\;{\rm cm}$ we have  $\Delta a \approx GM_\oplus/R^3 L \approx 2\cdot10^{-8}\;{\rm m}/{\rm s}^2$. In a rough estimate we assume this $\Delta a$ to act on the top surface of the cube. 
Now Hook's simple law of elasticity
\begin{equation}
\frac{F}{A} = E \frac{\Delta L}{L}
\end{equation}
gives the change of the length $\Delta L$ of the cube due to a force $F$ acting on the area $A$. In our case taking $F = m \Delta a = \rho L^3 \Delta a$ we get
\begin{equation}
\frac{\Delta L}{L} = \frac{\rho L}{E} \Delta a \approx 10^{-17}
\end{equation}
assuming an elasticity modulus of $E = 90\;{\rm GPa}$ and a density of $2350\;{\rm kg/m}^3$ which is typical for Zerodur.

In the OPTIS mission, for example, the science goal for the measurement of the isotropy of the speed of light is 
better than $\Delta c / c = 10^{-18}$ \cite{laemmerzahl04:_optis}. This can only be achieved if the resonator has 
a length stability of $\Delta L / L = 10^{-18}$  \cite{laemmerzahl04:_optis}. As one can see from our estimates, 
the tidal gravitational force will lead to systematic deformations which are one order of magnitude larger than the 
expected accuracy. Therefore the effect has to be investigated carefully by including 
the tidal gravitational force into the equations of elasticity, calculating the resulting resonator shape, 
and then subtract the effect. 

Although the linear theory of elasticity has a long history, explicit solutions for special problems are purely spread. In textbooks only examples for simple bodies in homogeneous gravitational fields or for thermal expansions can be found (e.g. \cite{LandauLifschitz}, \cite{Leipholz}, \cite{Lurje}). However, most of the solutions employ an ansatz which already includes knowledge about the expected solution. To the understanding of the authors, no publications are available dealing with a body under the influence of 
a tidal gravitational force so far. The reason for this is probably, that this situation applies only to bodies freely flying in space -- a situation which was outside the scope of application in elasticity theory so far. 

In the present paper we first derive an analytical solution in terms of a series expansion. This result is then confirmed using numerical methods. These calculations are usually done with help of Finite Element Method (FEM) codes. For most engineering purposes FEM codes are fine. However FEM solutions are only numerical approximations whose accuracy depends highly on the number and shape of the elements that have been chosen to mesh the model. In order to confirm the analytical model and to test the numerical calculation, we compare the analytical with the numerical solution. For this comparison we choose a cylinder as most simple geometry of a body adapted to the symmetry of the problem. 

Having thus checked the principal applicability of the FEM methods to these kinds of physical situations, this method safely can be used for calculating the deformations of arbitrarily shaped bodies or for the design of devices insensitive to unwanted influences, or for the elimination of the systematics of the measurements in order to ensure the success of highly sensitive experiments. 


\section{Basic Equations}

\subsection{Generalities}

The problem of an optical resonator flying on a geodetic Earth orbit can be simplified by treating the problem in 
a body fixed coordinate system. We also consider, for simplicity, the body to be a homogeneous and isotropic 
cylinder. The only force present is a volume force due to the tidal gravitational force which will be modeled 
as gradient of a spherically symmetric Earth acceleration field.

In order to calculate the elastic deformations of the cylinder the equations of elasticity have to be solved including 
the influence of the tidal gravitational force. The boundary conditions for the solution are given through the condition of weightlessness in space. 

As a short introduction, some basic equations of the linear theory of elasticity are given 
\cite{Leipholz,LandauLifschitz,Lurje,Kienzler}. All equations refer to homogeneous isotropic bodies.
Within this paper we do not use the notation within the formalism of the Riemannian geometry (e.g. \cite{Marsden}) but the notation used in \cite{Leipholz}. 

In elasticity the general relation between the stress tensor $\sigma_{ij}$ and the strain tensor $\eps_{ij}$ is given by Hooke's law
\begin{equation}
\sigma_{ij} = C_{ijkl}\eps_{ij} \qquad \qquad i,j,k,l = 1,2,3
\end{equation}
where $C_{ijkl}$ is the elasticity tensor related to the material under consideration. For homogeneous isotropic materials the elasticity tensor can be written as 
\begin{equation}
C_{ijkl} = \la \delta_{ij} \delta_{kl} + \mu (\delta_{ik}\delta_{jl}+\delta_{il}\delta_{jk}) \, ,
\end{equation}
where $\la$ and $\mu$ are the Lam{\'e} constants and $\delta_{mn}$ is the Kronecker symbol. Thus Hook's law for homogeneous isotropic materials is
\begin{equation}
\sigma_{ij} = \la \delta_{ij} \eps_{kk} + 2\mu \eps_{ij} \ .
\end{equation}

The strain tensor $\eps$ has to fulfill the so-called compatibility condition
\begin{equation}
\frac{\partial^2 \eps_{il}}{\partial r_j \partial r_k} 
+ \frac{\partial^2 \eps_{jk}}{\partial r_i \partial r_l} 
- \frac{\partial^2 \eps_{jl}}{\partial r_i \partial r_k}
- \frac{\partial^2 \eps_{ik}}{\partial r_j \partial r_l} = 0 \, ,
\end{equation}
where $r_n$ are the components of the position vector. The relations between strain and the displacement $\xi_i$ are
\begin{equation}
\eps_{ij} = \frac{1}{2} \left( \frac{\partial \xi_i}{\partial r_j} + \frac{\partial \xi_j}{\partial r_i}\right) \ .
\end{equation}

The equilibrium equation of elasticity describes the equilibrium state of a
homogeneous isotropic body when a volume force $\vec{K}$ is acting
\begin{equation}
\frac{\partial}{\partial r_j} \sigma_{ij} + K_i = 0 \ .
\end{equation}
Applying the relations between stress and displacements the equilibrium equation takes the form \cite{Leipholz}
\begin{equation}\label{Eq:ggw}
(\lambda+\mu) \frac{\partial^2}{\partial r_k \partial r_j} \xi_k + 
\mu \frac{\partial^2}{\partial r_i \partial r_i} \xi_j + K_j = 0 \, ,
\end{equation}
where $\vec{\xi}$  is the displacement vector. 
This equation can also be written as 
\begin{equation}\label{Eq:elast_displ}
\Delta \vec{\xi} + \frac{1}{1-2\nu} \nabla (\nabla \cdot \vec{\xi})
+ \frac{1}{\mu} \vec{K} = 0 
\end{equation}
where $\nu$ is the Poisson number which lies between 0 and 0.5 for homogeneous isotropic bodies. 

For vanishing volume forces $\vec{K}=0$ Eq.~(\ref{Eq:ggw}) becomes the homogeneous equilibrium equation
\begin{equation}
(\lambda+\mu) \frac{\partial^2}{\partial r_k \partial r_j} \xi_k + 
\mu \frac{\partial^2}{\partial r_i \partial r_i} \xi_j  = 0 \ .
\end{equation}

The boundary conditions for the solution of the equilibrium equation are either given by the forces 
$p_i$ acting on the body surfaces
\begin{equation}
\sigma_{ij} n_{j} = p_i  
\end{equation}
or by initial displacements $\xi_{i0}$ of the surfaces
\begin{equation}
\xi_i(0) = \xi_{i0} 
\end{equation}
where $n_j$ are the normal vectors on the surfaces.

The general solution of Eq.~(\ref{Eq:elast_displ}) is a
superposition of a homogeneous and a particular solution
\begin{equation}\label{Eq:xi_total}
\vec{\xi} = \vec{\xi}^{\rm h} + \vec{\xi}^{\rm p} \, .
\end{equation}

\subsection{The symmetries of our problem}

Since we have an axial symmetric problem, we use cylindrical coordinates $r, \vphi, z$ is useful. All displacements and derivatives with respect to
$\vphi$ vanish and the equilibrium equation of elasticity takes the form (see e.g. \cite{Leipholz})
\begin{eqnarray}
0 &=& \Delta \xi_r - \frac{\xi_r}{r^2} + \frac{1}{1-2\nu} \frac{\partial}{\partial r} 
\left( \frac{\partial \xi_r}{\partial r} + \frac{\xi_r}{r} + 
\frac{\partial \xi_z}{\partial z}\right) \nonumber \\
0 &=& \Delta \xi_z  + \frac{1}{1-2\nu} \frac{\partial}{\partial z} 
\left( \frac{\partial \xi_r}{\partial r} + \frac{\xi_r}{r} + 
\frac{\partial \xi_z}{\partial z}\right) \, .
\end{eqnarray}
The Laplace operator acting on a scalar takes the form
\begin{equation}\label{Eq:Delta_sym1}
\Delta = \frac{\partial^2}{\partial r^2} + \frac{1}{r}
\frac{\partial}{\partial r} + \frac{\partial^2}{\partial z^2} \, .
\end{equation}
Note that the Laplace operators acting on a vector field $\vec{\xi}$ takes the form (see \cite{Kovalenko})
\begin{eqnarray}\label{Eq:Delta_sym}
\Delta \xi_r &=& \frac{\partial^2 \xi_r}{\partial r^2} + \frac{1}{r}
\frac{\partial \xi_r}{\partial r} + \frac{\partial^2 \xi_r}{\partial z^2} -\frac{\xi_r}{r^2} \nonumber \\
\Delta \xi_z &=& \frac{\partial^2 \xi_z}{\partial r^2} + \frac{1}{r}
\frac{\partial \xi_z}{\partial r} + \frac{\partial^2 \xi_z}{\partial z^2} \ .
\end{eqnarray}

The relations between stresses, strains and displacements are
\begin{eqnarray}\label{Eq:stress_strain_displ}
\sigma_{rr} &=& \lambda (\eps_{rr} + \eps_{\vphi\vphi} + \eps_{zz} ) +
2 \mu \eps_{rr}
= \lambda \left( \frac{\partial \xi_r}{\partial r} + \frac{\xi_r}{r} + \frac{\partial \xi_z}{\partial z} \right) +
2 \mu \frac{\partial \xi_r}{\partial r} \nonumber \\ 
\sigma_{\vphi\vphi} &=& \lambda (\eps_{rr} + \eps_{\vphi\vphi} + \eps_{zz} ) +
2 \mu \eps_{\vphi\vphi}
= \lambda \left( \frac{\partial \xi_r}{\partial r} + \frac{\xi_r}{r} + \frac{\partial \xi_z}{\partial z} \right) +
2 \mu  \frac{\xi_r}{r} \nonumber \\
\sigma_{zz} &=& \lambda (\eps_{rr} + \eps_{\vphi\vphi} + \eps_{zz} ) +
2 \mu \eps_{zz}
= \lambda \left( \frac{\partial \xi_r}{\partial r} + \frac{\xi_r}{r} + \frac{\partial \xi_z}{\partial z} \right) +
2 \mu \frac{\partial \xi_z}{\partial z}  \nonumber \\
\sigma_{r \vphi} &=& 0 \nonumber \\
\sigma_{\vphi z} &=& 0 \nonumber \\
\sigma_{r z} &=& 2 \mu \eps_{rz}
=  \mu \left( \frac{\partial \xi_r}{\partial z} + \frac{\partial \xi_z}{\partial r} \right) \ .
\end{eqnarray}

Beside the axial symmetry we also have the following symmetries for reflection at the 
$z = 0$ plane: $\xi_z(r, -z) = - \xi_z(r, z)$ and $\xi_r(r, -z) = \xi_r(r, z)$.


\section{The Problem}\label{Sec:problem}

\begin{figure}
\centering\psfig{file=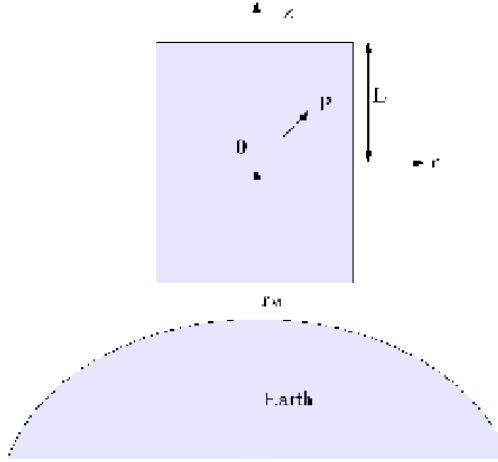,width=0.6\linewidth}
\caption{\label{Fig:gg_cylin}Simplified model of an optical resonator on a geodetic Earth orbit}
\end{figure}

In order to solve the problem of a free-flying isotropic homogeneous cylinder in space 
the equilibrium equation of elasticity (\ref{Eq:elast_displ}) has to be solved. The cylinder has radius $R$ and height $2L$. 
The body coordinates are $(r, \varphi, z )$ with the origin being at the center--of--mass of the cylinder. 
The $z$--axis coincides with the symmetry axis of the cylinder, see Fig.~\ref{Fig:gg_cylin}. 

The only force present is the volume force $\vec{K}$ which is due to the Earth's gravitational potential $U$, 
\begin{equation}\label{Eq:elast_displ_gradU}
\Delta \vec{\xi} + \frac{1}{1-2\nu} \nabla (\nabla \cdot \vec{\xi})
- \frac{1}{\mu} \rho \nabla U = 0  \, .
\end{equation}
For a spherical Earth potential, $U(r) = G M_\oplus/r$, where $GM_\oplus$ is the gravitational constant times the mass of Earth, 
the potential acting at an arbitrary point $P$ inside the cylinder can be calculated via Taylor expansion  
\begin{eqnarray}\label{Eq:Ugg}
U(\vec r_M + \vec r) &=& U(\vec r_M) + \frac{\partial U(\vec r_M)}{\partial r_i} r_i 
+ \frac{1}{2} \frac{\partial^2 U(\vec r_M)}{\partial r_i \partial r_j} r_i r_j \\
 &=& U(\vec r_M) + \nabla U(\vec r_M) \vec r + \frac{GM_\oplus}{2 r_{M}^3}(r^2-2z^2) 
\end{eqnarray}
where $\vec r_M$ is the vector from the center--of--mass of the Earth to the center--of--mass of the cylinder 
and $\vec r$ is the vector from the cylinder center--of--mass to point $P$. 
This Taylor expansion around the center--of--mass of the cylinder to second order gives the axis-symmetric potential in 
cylindrical coordinates. Note that the linear term of the Taylor expansion vanishes as this equation is valid in the freely 
falling reference frame of the cylinder. 

Since we consider a freely flying cylinder in an orbit around the Earth, no external forces are present and, thus, the forces $\vec p$ at the cylinder surfaces are zero which gives us the boundary conditions 
\begin{equation}\label{Eq:boundary}
\sigma_{ij} n_{j} = p_i = 0 \ .
\end{equation}

The normal vector $\vec{n} = (n_r \, n_{\vphi} \, n_z)^T$ ($T$ means the transposed vector) reduces in the axis-symmetric case to $\vec{n} = (n_r \, 0 \, n_z)^T$, as the $\vphi$ component is zero. 

Thus the boundary conditions (\ref{Eq:boundary}) at the top surface of the cylinder, i.e. for $z=L$, $\vec{n} = (0 \, 0 \, 1)^T$, are 
\begin{eqnarray}
p_r(r,z=L) &=& 0 = \sigma_{rz}(r, z=L) \nonumber \\
p_z(r,z=L) &=& 0 = \sigma_{zz}(r, z=L) \ .
\end{eqnarray}
At the bottom surface of the cylinder, i.e. for $z=-L$, $\vec{n} = (0 \, 0 \, {-1})^T$, we have
\begin{eqnarray}
p_r(r,z=-L) &=& 0 = -\sigma_{rz}(r, z=-L) \nonumber \\
p_z(r,z=-L) &=& 0 = -\sigma_{zz}(r, z=-L)  \ .
\end{eqnarray}
Note that these boundary conditions are valid for arbitrary $r \in [0,R)$.
For $r=R$ they are not valid as the normal vector is not uniquely defined at the cylinder 
edges $(r,z)=(R,\pm L)$.

For the superficies cylinder surface, i.e. $r=R$, $\vec{n} = (1 \, 0 \,  0)^T$, 
the boundary conditions are 
\begin{eqnarray}
p_r(r=R,z) &=& 0 = \sigma_{rr}(r=R, z) \nonumber \\
p_z(r=R,z) &=& 0 = \sigma_{zr}(r=R, z) 
\end{eqnarray}
for all $z \in [0,\pm L)$.


\section{The Solution}

\subsection{Particular Solution}\label{Sec:part_sol}

In order to find a particular solution of the problem one can assume that the 
solution of the equilibrium equation can be written as gradient of a scalar $\psi$
\cite{Lurje} 
\begin{equation}
\vec{\xi}^{\rm p} = \nabla \psi \, .
\end{equation}
Inserting this approach into Eq.~(\ref{Eq:elast_displ_gradU}) yields
\begin{eqnarray}
\nabla \left(\Delta \psi + \frac{1}{1-2\nu}\Delta\psi \right) &=& 
\frac{\rho}{\mu} \nabla U \nonumber \\
\nabla \left( \frac{2(1-\nu)}{1-2\nu} \Delta \psi\right) &=& 
\frac{\rho}{\mu} \nabla U \ .
\end{eqnarray}
Calculation of the volume integral of the divergence gives
\begin{equation}\label{Eq:Dpsi}
\Delta \psi = \frac{1-2\nu}{2(1-\nu)\mu} \rho (U+U_0) 
\end{equation}
where $U_0$ is the integration constant. Inserting the potential $U$ calculated in Eq.~(\ref{Eq:Ugg}) 
and using (\ref{Eq:Delta_sym1}) we obtain  
\begin{equation}
\psi = -\frac{1-2\nu}{2(1-\nu)\mu} \rho
\frac{GM_\oplus}{2 r_{M}^3}\left(-\frac{r^4}{16}+\frac{1}{6} z^4 + cr^2 + dz^2
+c_1 r + d_1z +c_2 +d_2\right)
\end{equation}
where $U_0=4c+2d$.
Thus, the displacement vector becomes
\begin{equation}
\vec{\xi}^{\rm p} = -\frac{1-2\nu}{2(1-\nu)\mu} \rho
\frac{GM_\oplus}{2 r_{M}^3}\left( \left(-\frac{r^3}{4}+2cr + c_1\right) \vec{e_r} + 
\left( \frac{2}{3} z^3 +2dz + d_1\right) \vec{e_z} \right) \, .
\end{equation} 
Since the displacement at the center of mass of the cylinder should vanish, $\xi_r(0,0)=\xi_z(0,0)=0$, both, $c_1$ and $d_1$, have to be zero. The unknown constants $c$ and $d$ are arbitrary. 

The $r$ and $z$ components of the displacement are (with $\mu=\lambda \frac{1-2\nu}{2\nu}$)
\begin{eqnarray}
\xi_r^{\rm p} &=& -\frac{1-2\nu}{2(1-\nu)\mu} \rho
\frac{GM_\oplus}{2 r_{M}^3} \left(-\frac{r^3}{4}+2cr\right)
=\frac{\nu}{\lambda(1-\nu)}\gamma \left(-\frac{r^3}{4}+2cr\right) \\ \nonumber
\xi_z^{\rm p} &=& -\frac{1-2\nu}{2(1-\nu)\mu} \rho
\frac{GM_\oplus}{2 r_{M}^3} \left(\frac{2}{3} z^3 + 2dz\right)
=\frac{\nu}{\lambda(1-\nu)}\gamma \left(\frac{2}{3} z^3 + 2dz\right) \,  ,
\end{eqnarray}
where we substituted $\gamma: = - G M/(2 r_{M}^3) \rho$.

Using Eqs.~(\ref{Eq:stress_strain_displ}) we obtain the stress components out of the displacements
\begin{eqnarray}\label{Eq:sigma_p}
\sigma_{rr}^{\rm p} &=& \gamma \left( \frac{(2\nu-3)}{4(1-\nu)}r^2 +\frac{2(c+\nu d)}{1-\nu}+\frac{2\nu }{1-\nu}z^2\right)\nonumber \\
\sigma_{zz}^{\rm p} &=& \gamma \left( \frac{4c\nu}{1-\nu} -\frac{\nu}{1-\nu}r^2  +2(z^2+d)\right) \nonumber \\
\sigma_{rz}^{\rm p} &=& 0  \ .
\end{eqnarray}


\subsection{Homogeneous Solution}

\subsubsection{Derivation of the boundary conditions}

A displacement vector which satisfies the homogeneous equation of elasticity 
\begin{equation}\label{Eq:elast_displ_hom}
\Delta \vec{\xi}^{\rm h} + \frac{1}{1-2\nu} \nabla (\nabla \cdot \vec{\xi}^{\rm h}) = 0 
\end{equation}
has to fulfill the biharmonic equation (see, e.g. \cite{Leipholz} and \cite{Boresi})
\begin{equation}\label{Eq:biharm}
\nabla^2 \nabla^2 \vec{\xi}^{\rm h} = \Delta \Delta \vec{\xi}^{\rm h} = 0 \ .
\end{equation}
Note that the Laplace operator applied to a vector field has the form given in Equation (\ref{Eq:Delta_sym}).

The boundary conditions the homogeneous solution part has to fulfill can be
derived from the boundary conditions (\ref{Eq:boundary}) of the complete solution. 
The boundary conditions at the cylinder top surface, i.e. $z=L$, 
normal vector $\vec{n} = (0 \, 0\, 1)^T$, are
\begin{eqnarray}\label{Eq:boundary_top}
p_r(r,z=L) &=& 0 = \sigma_{rz}(r, z=L) =  
\sigma_{rz}^{\rm h}(r, z=L) + \sigma_{rz}^{\rm p}(r, z=L) \nonumber \\
p_z(r,z=L) &=& 0 = \sigma_{zz}(r, z=L) =  \sigma_{zz}^{\rm h}(r, z=L) + \sigma_{zz}^{\rm p}(r, z=L) \ .
\end{eqnarray}
With help of the expressions of the stress components derived from the particular solution (\ref{Eq:sigma_p}) one obtains
\begin{eqnarray}\label{Eq:b_posL}
\sigma_{zz}^{\rm h}(r, z=L)&=& -\sigma_{zz}^{\rm p}(r, z=L) =
-\gamma \left( \frac{4c\nu}{1-\nu} -\frac{\nu}{1-\nu}r^2  +2(L^2+d)\right)\nonumber \\
\sigma_{rz}^{\rm h}(r, z=L) &=&  -\sigma_{rz}^{\rm p}(r, z=L)= 0 \, .
\end{eqnarray}

The boundary conditions at the cylinder bottom surface, i.e. $z=-L$, 
normal vector, $\vec{n} = (0 \, 0\, {-1})^T$
\begin{eqnarray}\label{Eq:boundary_bottom}
p_r(r,z=-L) &=& 0 = -\sigma_{rz}(r, z=-L) \nonumber \\
 &=&  -\sigma_{rz}^{\rm h}(r, z=-L) - \sigma_{rz}^{\rm p}(r, z=-L) \nonumber \\
p_z(r,z=-L) &=& 0 = -\sigma_{zz}(r, z=-L) \nonumber \\
&=&  -\sigma_{zz}^{\rm h}(r, z=-L) - \sigma_{zz}^{\rm p}(r, z=-L)  
\end{eqnarray}
give the same boundary conditions as for the top surface, as
$z$ occurs only as $z^2$ term
\begin{eqnarray}\label{Eq:b_negL}
\sigma_{zz}^{\rm h}(r, z=-L) &=& -\sigma_{zz}^{\rm p}(r, z=-L) \nonumber \\ 
&=&-\gamma \left( \frac{4c\nu}{1-\nu} -\frac{\nu}{1-\nu}r^2  +2(L^2+d)\right)\nonumber \\
\sigma_{rz}^{\rm h}(r, z=-L) &=&  -\sigma_{rz}^{\rm p}(r, z=-L) = 0 \qquad .
\end{eqnarray}

For the superficies surface, i.e. $r=R$, $\vec{n} = (1 \,0\, 0)^T$
\begin{eqnarray}\label{Eq:boundary_r}
p_r(r=R, z) &=& 0 = \sigma_{rr}(r=R, z) =  
\sigma_{rr}^{\rm h}(r=R, z) + \sigma_{rr}^{\rm p}(r=R, z) \nonumber \\
p_z(r=R, z) &=& 0 = \sigma_{zr}(r=R, z) =  \sigma_{zr}^{\rm h}(r=R, z) + \sigma_{zr}^{\rm p}(r=R, z)  
\end{eqnarray}
one gets
\begin{eqnarray}\label{Eq:b_R}
\sigma_{rr}^{\rm h}(r=R, z) &=& -\sigma_{rr}^{\rm p}(r=R, z) \nonumber \\ 
&=&-\gamma \left( \frac{(2\nu-3)}{4(1-\nu)}R^2 +\frac{2(c+\nu d)}{1-\nu}+\frac{2\nu }{1-\nu}z^2\right)\nonumber \\
\sigma_{zr}^{\rm h}(r=R, z) &=&  -\sigma_{zr}^{\rm p}(r=R, z) = 0 \qquad .
\end{eqnarray}


\subsubsection{General Ansatz for the Homogeneous Solution}

Love \cite{Love} showed, that the deformations in an elastic axis-symmetric body 
can be expressed in terms of the so-called Love function $\chi$
\begin{eqnarray}\label{Eq:xiLove}
\xi_r^{\rm h} &=& -\frac{1+\nu}{E} \frac{\partial^2 \chi}{\partial r \partial z} \nonumber \\
\xi_z^{\rm h} &=& \frac{1+\nu}{E} \left( (1-2\nu) \nabla^2 \chi + \frac{\partial^2 \chi}{\partial r^2} + \frac{1}{r} \frac{\partial \chi}{\partial r} \right)
\end{eqnarray}
where $\xi_r$ and $\xi_z$ are the displacement components.

The corresponding stress components are
\begin{eqnarray}\label{Eq:sigmaLove}
\sigma_{rr}^{\rm h} &=& \frac{\partial}{\partial z} \left( \nu \nabla^2 \chi - \frac{\partial^2 \chi}{\partial r^2} \right) \nonumber \\
\sigma_{rz}^{\rm h} &=& \frac{\partial}{\partial r} \left( (1-\nu) \nabla^2 \chi - \frac{\partial^2 \chi}{\partial z^2} \right) \nonumber \\
\sigma_{zz}^{\rm h} &=& \frac{\partial}{\partial z} \left( (2-\nu) \nabla^2 \chi - \frac{\partial^2 \chi}{\partial z^2} \right) \nonumber \\
\sigma_{\phi\phi}^{\rm h} &=& \frac{\partial}{\partial r} \left( \nu \nabla^2 \chi - \frac{1}{r} \frac{\partial \chi}{\partial r} \right) \ .
\end{eqnarray}
The Love function $\chi$ necessarily fulfills the biharmonic equation
\begin{equation}
\nabla^2 \nabla^2 \chi = 0 \ .
\end{equation}

The major obstacle is to find an adequate approach for the Love function fulfilling all boundary conditions. A general approach is the separation of variables, that means a factorization of the kind $\chi(r,z)=F_1(r)\cdot F_2(z)$. For axial symmetric problems the Bessel functions (see \ref{A:bessel}) are the natural choice to  represent the $r$ dependence. Owing to the additional reflection symmetry and antisymmetry of $\xi_r$ and $\xi_z$, the $z$ dependence can be represented by the trigonometric functions cosine and sine. Bessel functions as well as trigonometric functions form a complete orthogonal set of functions so that each function can be represented as a series of these sets. 

For the solution of the homogeneous equation of elasticity a 
so-called Papkovich-Neuber approach for the displacement field can be made  \cite{Kovalenko}. One writes  $\vec{\xi} = \nabla \Phi + 4 (1 - \nu) \vec{B} - \nabla \left(\vec{B} \cdot \vec{r} + B_0\right)$, where $\Phi$ is a scalar, and the vector $\vec{B}$ fulfills the biharmonic equation. For $\vec{B}$ and $B_0$ the approach of separation of variables is used where the $r$ dependency is represented by Bessel functions and the $z$ dependency is represented by trigonometric functions. 
By modifying this approach one can make an ansatz for the Love function suggested in \cite{Meleshko}
\begin{eqnarray}\label{Eq:chi}
\chi &=& B_0 z^3 + \sum_{j=1}^{\infty} \left( A_j \frac{\sinh(\la_j z)}{\sinh(\la_j L)} + B_j z \frac{\cosh(\la_j z)}{\sinh(\la_j L)}\right) \frac{J_0(\la_j r)}{\la_j^2} \nonumber \\
&&+ D_0 r^2 z+ \sum_{n=1}^{\infty} \left( C_n \frac{I_0(k_nr)}{I_1(k_nR)} + D_n r \frac{I_1(k_nr)}{I_1(k_nR)}\right) \frac{\sin(k_nz)}{k_n^2} \ .
\end{eqnarray} 
Herein $J_0$ are the Bessel functions of first kind and order zero and $I_1$ are the modified
Bessel functions of order one (see \ref{A:bessel}). 
Furthermore $\zeta_j=\la_jR$ are the zeros of the Bessel functions of order one, $J_1(\zeta_j)=0$, and $k_n = \frac{n\pi}{L}$ where $n$ is an integer number. 

If we insert the Love function approach into (\ref{Eq:sigmaLove}) 
we obtain for the stress components
\begin{eqnarray}\label{Eq:sigmaterms}
\sigma_{rr}^{\rm h} &=& 6\nu B_0+(4\nu-2)D_0 \\
&&+ \sum_{j=1}^{\infty} \left( A_j \la_j \frac{\cosh(\la_j z)}{\sinh(\la_j L)} + B_j \left( (1+2\nu)\frac{\cosh(\la_j z)}{\sinh(\la_j L)} + \la_jz \frac{\sinh(\la_j z)}{\sinh(\la_j L)}\right) \right) J_0(\la_j r) \nonumber \\ 
&&- \sum_{j=1}^{\infty} \left( A_j \la_j \frac{\cosh(\la_j z)}{\sinh(\la_j L)} + B_j \left( \frac{\cosh(\la_j z)}{\sinh(\la_j L)} + \la_jz \frac{\sinh(\la_j z)}{\sinh(\la_j L)}\right) \right) \frac{J_1(\la_j r)}{\la_jr} \nonumber \\
&&- \sum_{n=1}^{\infty} \left(C_n k_n\left( \frac{I_0(k_nr)}{I_1(k_nR)}-\frac{1}{k_nr}\frac{I_1(k_nr)}{I_1(k_nR)} \right) \right.\nonumber \\
&&\left.+ D_n \left( k_nr \frac{I_1(k_nr)}{I_1(k_nR)} + (1-2\nu) \frac{I_0(k_nr)}{I_1(k_nR)}\right) \right) \cos(k_nz) \nonumber \\
\sigma_{zz}^{\rm h} &=& (6-6\nu) B_0+(8-4\nu)D_0 \nonumber \\
&&- \sum_{j=1}^{\infty} \left( A_j \la_j \frac{\cosh(\la_j z)}{\sinh(\la_j L)} + B_j \left( (2\nu-1)\frac{\cosh(\la_j z)}{\sinh(\la_j L)} + \la_jz \frac{\sinh(\la_j z)}{\sinh(\la_j L)}\right) \right) J_0(\la_j r) \nonumber \\ 
&&+ \sum_{n=1}^{\infty} \left( C_n k_n \frac{I_0(k_nr)}{I_1(k_nR)}
+ D_n \left( (4-2\nu) \frac{I_0(k_nr)}{I_1(k_nR)} + k_nr \frac{I_1(k_nr)}{I_1(k_nR)}\right) \right) \cos(k_nz) \nonumber \\
\sigma_{rz}^{\rm h} &=& \sum_{j=1}^{\infty} \left( A_j \la_j \frac{\sinh(\la_j z)}{\sinh(\la_j L)} + B_j \left( 2\nu\frac{\sinh(\la_j z)}{\sinh(\la_j L)} + \la_jz \frac{\cosh(\la_j z)}{\sinh(\la_j L)}\right) \right) J_1(\la_j r) \nonumber \\ 
&&+\sum_{n=1}^{\infty} \left( C_n k_n \frac{I_1(k_nr)}{I_1(k_nR)}
+ D_n \left( (2-2\nu) \frac{I_1(k_nr)}{I_1(k_nR)} + k_nr \frac{I_0(k_nr)}{I_1(k_nR)}\right) \right) \sin(k_nz)\nonumber \ .
\end{eqnarray}   
The appearance of the coefficients $A_j$ and $B_j$ in the Dini and Bessel-Fourier series ($d_0+\sum_{n=1}^{\infty} d_j J_0(\la_jr)$ and $\sum_{n=1}^{\infty} c_j J_1(\la_jr)$, see \ref{A:bessel}) allows us to represent arbitrary boundary conditions for $\sigma_{zz}$ and $\sigma_{rz}$ at the cylinder top and bottom surfaces ($z=\pm L$). Similarly, the appearance of the coefficients $C_n$ and $D_n$ in the Fourier series ($a_0/2+\sum_{n=1}^{\infty}a_n\cos(k_nz)+\sum_{n=1}^{\infty}b_n\sin(k_nz)$, 
see \ref{B:fourier}) allows us to describe arbitrary boundary conditions for $\sigma_{rr}$ and $\sigma_{rz}$ at the superficies surface of the cylinder. By using  the boundary conditions we can now determine the unknown coefficients.


\subsubsection{Determination of the Coefficients $A_j$ and $C_n$}

For the boundary conditions $0=\sigma_{rz}^{\rm h}(r, z=\pm L)$ (Eqs.~(\ref{Eq:b_posL}) and (\ref{Eq:b_negL})) we get
\begin{eqnarray}
0 &=& \sum_{j=1}^{\infty} \left( A_j \la_j  + B_j \left( 2\nu + \la_jL \coth(\la_j L)\right) \right) J_1(\la_j r) \ .
\end{eqnarray}
Multiplication with $kJ_1(\la_{j'}r)rdr$ and subsequent integration over the interval $0$ to $R$ 
yields with help of (\ref{Eq:A_thorne41}) 
\begin{equation}\label{Eq:AjBj}
A_j \la_j  = - B_j \left( 2\nu + \la_jL \coth(\la_j L)\right) \qquad j=1\dots\infty \ .
\end{equation}

For the boundary conditions $0=\sigma_{rz}^{\rm h}(r=R,z)$ (Eq.~\ref{Eq:boundary_r}) we get
\begin{eqnarray}
0&=&\sum_{n=1}^{\infty} \left( C_n k_n + D_n \left( (2-2\nu) + k_nR \frac{I_0(k_nr)}{I_1(k_nR)}\right) \right) \sin(k_nz) \ .
\end{eqnarray}
Multiplication with $\sin(k_{n'}z)dz$ and subsequent integration over the interval $-L$ to $L$
yields with help of (\ref{B:ortho}) 
\begin{equation}\label{Eq:CnDn}
C_n k_n = -  D_n \left( (2-2\nu) + k_nR \frac{I_0(k_nR)}{I_1(k_nR)} \right) \qquad n=1\dots\infty \ .
\end{equation}


\subsubsection{Determination of the Coefficients $B_j$ and $D_n$}

A simpler expression for the boundary conditions can be obtained by inserting (\ref{Eq:AjBj}) and (\ref{Eq:CnDn}) into (\ref{Eq:sigmaterms}),  
\begin{eqnarray}
\sigma_{rr}^{\rm h}(r=R,z) &=&  -\sigma_{rr}^{\rm p}(r=R,z) \nonumber\\
&=& 6\nu B_0+(4\nu-2)D_0 \nonumber \\ 
&&+ \sum_{j=1}^{\infty} B_j \left( (1-\la_j L \coth(\la_j L)) \frac{\cosh(\la_j z)}{\sinh(\la_j L)} + \la_jz \frac{\sinh(\la_j z)}{\sinh(\la_j L)}\right) J_0(\la_j R) \nonumber\\
&&+ \sum_{n=1}^{\infty} D_n \left( k_nR \left( \frac{I_0(k_nR)^2}{I_1(k_nR)^2}-1\right) - \frac{2-2\nu}{k_nR} \right) \cos(k_nz)  \label{Eq:sig_b_r1} \\
\sigma_{zz}^{\rm h}(r,z=\pm L)  &=&  -\sigma_{zz}^{\rm p}(r,z=\pm L) \nonumber \\
&=& (6-6\nu) B_0+(8-4\nu)D_0 \nonumber\\
&&+ \sum_{j=1}^{\infty} B_j \left( \coth(\la_j L) + \frac{\la_j L}{\sinh^2(\la_j L)} \right) J_0(\la_j r) \nonumber\\ 
&&+ \sum_{n=1}^{\infty} (-1)^n D_n \left( k_nr \frac{I_1(k_nr)}{I_1(k_nR)} + \left(2-k_nR\frac{I_0(k_nR)}{I_1(k_nR)}\right) \frac{I_0(k_nr)}{I_1(k_nR)} \right)  \ . \label{Eq:sig_b_z1}
\end{eqnarray}
Again $J_1(\la_jR)=0$ and $\cos(k_nL)=\cos(n\pi)=(-1)^n$ were used.


\paragraph{Expansion of the Particular Boundary Conditions in Dini and Fourier Series}

As the Bessel functions form a complete orthogonal set of functions, each function can be represented by a Bessel series as given in (\ref{Eq:A_arfken1151}).
With the additional condition that $\zeta_j=\la_jR$ are the zeros of $J_1$ and
$\frac{\partial}{\partial r} J_0(\la_jr)|_{r=R} = -\la_j
J_1(\la_jr)|_{r=R} = -\la_j J_1(\la_jR) = -\la_jRJ_1(\zeta_j)=0$,
the particular boundary condition part $\sigma_{zz}^{\rm p}(r,z=\pm L)$ can be represented by
a Dini series expansion as given in Equation (\ref{Eq:A_dini})
\begin{equation}
\sigma_{zz}^{\rm p}(r,z=\pm L) = \sigma_{zz}^{\rm p}(r) = d_0 + \sum_{j=1}^{\infty} d_j J_0(\la_jr) 
\end{equation}
with 
\begin{equation}
d_0 = \frac{2}{R^2} \int_0^R  \sigma_{zz}^{\rm p}(r) r dr 
= \hat{c}_0-\frac{\hat{c}_2R^2}{4}
\end{equation}
where $\hat{c}_0=\gamma \left(\frac{4c\nu}{1-\nu}+2(L^2+d)\right)$ and 
$\hat{c}_2= 2 \gamma \frac{\nu}{1-\nu}$ were substituted
and
\begin{eqnarray}
d_j &=& \frac{2}{R^2[J_0(\zeta_j)]^2} 
\int_0^R  \sigma_{zz}^{\rm p}(r) J_0(\la_jr) r dr \\
    &=&  -\frac{4}{\la_j^2 J_0(\zeta_j)} \gamma \frac{\nu}{1-\nu} \qquad j=1\dots\infty \ .
\end{eqnarray}
Note that, following (\ref{Eq:A_thorne43}), $\int_0^R r J_0(\la_j r) r dr =0$ if
$\zeta_j=\la_jR$ are the zeros of $J_1$. 

Therefore the particular boundary condition $\sigma_{zz}^{\rm p}(r,z=\pm L)$ can be written as
\begin{equation}
\sigma_{zz}^{\rm p}(r) = \hat{c}_0-\frac{\hat{c}_2R^2}{4} 
- \sum_{j=1}^{\infty} \frac{2}{\la_j^2 J_0(\zeta_j)} \hat{c}_2 J_0(\la_jr) \ .
\end{equation}

Similarly, each function can be represented by a Fourier series, Eq.~(\ref{B:Fseries})
\begin{equation}
\sigma_{rr}^{\rm p}(r=R,z) = \sigma_{rr}^{\rm p}(z) = \frac{a_0}{2}+\sum_{n=1}^{\infty} a_n \cos(k_n z) + \sum_{n=1}^{\infty} b_n \sin(k_n z) 
\end{equation}
and the Fourier coefficients can be determined from (\ref{B:Fcoeff}). This gives a new representation of the particular boundary condition $\sigma_{rr}^{\rm p}(r=R,z)$
\begin{equation}
\sigma_{rr}^{\rm p}(z) = \frac{a_0}{2}+\sum_{n=1}^{\infty} a_n \cos(k_n z) 
= \hat{c}_1 +\frac{L^2 \hat{c}_2}{3} 
+ \frac{4L^2 \hat{c}_2}{\pi^2} \sum_{n=1}^{\infty}\frac{(-1)^n}{n^2}
\cos(k_n z) \, ,
\end{equation}
where
\begin{equation}
\hat{c}_1 = \gamma \left( \frac{(2\nu-3)}{4(1-\nu)}R^2 +\frac{2(c+\nu
d)}{1-\nu}\right) 
\qquad {\rm and} \qquad
\hat{c}_2 = \gamma \frac{2\nu }{1-\nu}
\end{equation}
with the arbitrary constants $c$ and $d$ from the particular solution part.

\paragraph{Inserting the Particular Boundary Conditions}

After having found representations for the particular boundary conditions in terms of cos and $J_0$ we can determine the remaining unknown coefficients from Eqs.~(\ref{Eq:sig_b_r1}) and (\ref{Eq:sig_b_z1}). In doing so we first simplify these equations with help of (\ref{B:A1}), (\ref{B:A4}) and (\ref{B:A5})
\begin{eqnarray}\label{Eq:sig_b_rr}
-\sigma_{rr}^{\rm p}(r=R,z) &=&  \sigma_{rr}^{\rm h}(r=R,z)  \\
 -\hat{c}_1 -\frac{L^2 \hat{c}_2}{3}- \frac{4L^2 \hat{c}_2}{\pi^2} \sum_{n=1}^{\infty}\frac{(-1)^n}{n^2}
\cos(k_n z) &=& 6\nu B_0+(4\nu-2)D_0 \nonumber \\
&&+ \sum_{j=1}^{\infty} B_j \sum_{n=1}^{\infty} (-1)^n \frac{4\la_j k_n^2}{L(k_n^2+\la_j^2)^2} \cos(k_n z) J_0(\la_j R) \nonumber \\
&&+ \sum_{n=1}^{\infty} D_n \left( k_nR \left( \frac{I_0(k_nR)^2}{I_1(k_nR)^2}-1\right) - \frac{2-2\nu}{k_nR} \right) \cos(k_nz) \nonumber
\end{eqnarray}
\begin{eqnarray}\label{Eq:sig_b_zz}
-\sigma_{zz}^{\rm p}(r,z=\pm L)  &=&  \sigma_{zz}^{\rm h}(r,z=\pm L) \\
-\hat{c}_0+\frac{\hat{c}_2R^2}{4}+\sum_{j=1}^{\infty} \frac{2}{\la_j^2 J_0(\zeta_j)} \hat{c}_2 J_0(\la_jr)
 &=& (6-6\nu) B_0+(8-4\nu)D_0 \nonumber \\
&&+ \sum_{j=1}^{\infty} B_j \left( \coth(\la_j L) + \frac{\la_j L}{\sinh^2(\la_j L)} \right) J_0(\la_j r) \nonumber \\
&&+\sum_{n=1}^{\infty} (-1)^n D_n k_n \left( \sum_{j=1}^{\infty} \frac{4\la_j^2}{R(k_n^2+\la_j^2)^2} \frac{J_0(\la_j r)}{J_0(\la_j R)} \right) \ . \nonumber
\end{eqnarray} 

Multiplication of (\ref{Eq:sig_b_rr}) with $\cos(k_{n'}z)$ and subsequent integration over the interval from $-L$ to $L$ gives with help of (\ref{B:ortho})
\begin{equation}
-\frac{4L^2 \hat{c}_2}{\pi^2 n^2} = \sum_{j=1}^{\infty} B_j \frac{4\la_j k_n^2}{L(k_n^2+\la_j^2)^2} J_0(\la_j R) 
+ (-1)^n  D_n \left( k_nR \left( \frac{I_0(k_nR)^2}{I_1(k_nR)^2}-1\right) - \frac{2-2\nu}{k_nR} \right)  \ .
\end{equation}
This results in an equation for the coefficients $D_n$ in dependence of $B_n$
\begin{eqnarray}\label{Eq:DnBj_1}
D_n & = & - \left( k_nR \left(
\frac{I_0(k_nR)^2}{I_1(k_nR)^2}-1\right) - \frac{2-2\nu}{k_nR} \right)^{-1} \times \nonumber\\
& & \qquad \times \left(\sum_{m=1}^{\infty} B_m (-1)^n \frac{4\la_m
k_n^2}{L(k_n^2+\la_m^2)^2} J_0(\la_m R) + 4 \frac{(-1)^n \hat{c}_2}{k_n^2}\right) \ .
\end{eqnarray}

In the same way multiplication of (\ref{Eq:sig_b_zz}) with $J_0(\la_{j'}r)r$ 
and subsequent integration over the interval from $0$ to $R$ gives with help of  (\ref{Eq:A_thorne42})
\begin{eqnarray}
\frac{2}{\la_j^2} \hat{c}_2 & = & J_0(\la_jR) B_j \left( \coth(\la_j L) + \frac{\la_j
L}{\sinh^2(\la_j L)} \right) \nonumber\\
& & \qquad 
+\sum_{n=1}^{\infty} (-1)^n D_n k_n \left( 
\frac{4\la_j^2}{R(k_n^2+\la_j^2)^2}  \right) \ ,
\end{eqnarray}
what gives an equation for the coefficients $B_j$ in dependence of $D_n$
\begin{eqnarray}\label{Eq:DnBj_2}
B_j & = & \left(J_0(\la_jR) \left( \coth(\la_j L) + \frac{\la_j
L}{\sinh^2(\la_j L)} \right)\right)^{-1} \times \nonumber\\
& & \qquad \times \left(\frac{2}{\la_j^2} \hat{c}_2 - \sum_{k=1}^{\infty} (-1)^k D_k k_k \left(\frac{4\la_j^2}{R(k_k^2+\la_j^2)^2}  \right) \right) \ . 
\end{eqnarray}

\subsubsection{Determination of $B_0$ and $D_0$}

After the determination of the coefficients in the infinite series (\ref{Eq:chi}) 
we determine the remaining unknowns $B_0$ and $D_0$. From the boundary condition $\sigma_{rr}^{\rm h}(r=R,z)=-\sigma_{rr}^{\rm p}(r=R,z)$, valid for all $z \in [0,\pm L)$, one can deduce that in particular $\sigma_{rr}^{\rm h}(r=R,z=0)=-\sigma_{rr}^{\rm p}(r=R,z=0)$ must hold. Then Eq.~(\ref{Eq:sig_b_r1}) simplifies to
\begin{eqnarray}
 -\hat{c}_1 -\frac{L^2 \hat{c}_2}{3}- \frac{4L^2 \hat{c}_2}{\pi^2} \sum_{n=1}^{\infty}\frac{(-1)^n}{n^2} 
&=& 6\nu B_0+(4\nu-2)D_0 \nonumber \\ 
&&+ \sum_{j=1}^{\infty} B_j \left( \frac{(1-\la_j L \coth(\la_j L))}{\sinh(\la_j L)}\right) J_0(\la_j R) \nonumber \\ 
&&+ \sum_{n=1}^{\infty} D_n \left( k_nR \left( \frac{I_0(k_nR)^2}{I_1(k_nR)^2}-1\right) - \frac{2-2\nu}{k_nR} \right) \ . \nonumber\\ 
\end{eqnarray}
From $\sum_{n=1}^{\infty}\frac{(-1)^n}{n^2} = -\frac{\pi^2}{12}$ we get
\begin{equation}\label{Eq:RL_r}
-\hat{c}_1 = 6\nu B_0+(4\nu-2)D_0  + {\cal R} \, ,
\end{equation}
where we abbreviated
\begin{eqnarray}
{\cal R} &=& \sum_{j=1}^{\infty} B_j \left( \frac{(1-\la_j L \coth(\la_j L))}{\sinh(\la_j L)}\right) J_0(\la_j R) \nonumber\\
& & \qquad + \sum_{n=1}^{\infty} D_n \left(k_n R  \left(\frac{I_0(k_nR)^2}{I_1(k_nR)^2}-1\right) - \frac{2-2\nu}{k_nR} \right) \, .
\end{eqnarray} 

Similarly, from the boundary condition $\sigma_{zz}^{\rm h}(r,z=\pm L)=-\sigma_{zz}^{\rm p}(r,z=\pm L)$ valid 
for all $r\in [0,\pm R)$, one can deduce that in particular
$\sigma_{zz}^{\rm h}(r=0,z=L)=-\sigma_{zz}^{\rm p}(r=0,z=L)$ 
must be valid. 
Then Eq.~(\ref{Eq:sig_b_z1}) simplifies to
\begin{equation}
-\hat{c}_0 + \frac{R^2}{4}\hat{c}_2 + \sum_{j=1}^{\infty} \frac{2}{\la_j^2 J_0(\zeta_j)} \hat{c}_2 = (6-6\nu) B_0+(8-4\nu)D_0 + {\cal Z} \, ,
\end{equation}
where we abbreviated
\begin{eqnarray}
{\cal Z} & = & \sum_{j=1}^{\infty} B_j \left( \coth(\la_j L) + \frac{\la_j L}{\sinh^2(\la_j L)} \right)  \nonumber\\
& & \qquad 
+ \sum_{n=1}^{\infty} (-1)^n D_n \left( 2-k_nR\frac{I_0(k_nR)}{I_1(k_nR)}\right) \frac{1}{I_1(k_nR)} \ .
\end{eqnarray}
With $\sum_{j=1}^{\infty} \frac{1}{J_0(\zeta_j)\la_j^2} = -\frac{R^2}{8}$ (compare Eq.~(\ref{B:A11_var})) we obtain
\begin{equation}\label{Eq:RL_z}
-\hat{c}_0 = (6-6\nu) B_0+(8-4\nu)D_0 + {\cal Z} \ .
\end{equation}

Addition of Eqs.~(\ref{Eq:RL_r}) and (\ref{Eq:RL_z}) gives
\begin{equation}\label{Eq:B0D0}
-\hat{c}_0 -\hat{c}_1 = 6(B_0+D_0) + {\cal S}
\end{equation}
with ${\cal S} = {\cal R}+{\cal Z}$ from which we can determine $B_0$ and $D_0$. Eq.~ (\ref{Eq:B0D0}) yields
\begin{equation}\label{Eq:B0}
B_0 = - \frac{1}{6} \left({\hat{c}}_0 + {\hat{c}}_1 + {\cal S}\right) - D_0 \ .
\end{equation}
and by inserting (\ref{Eq:B0}) into (\ref{Eq:RL_r}) we get
\begin{equation}\label{Eq:D0}
D_0 =
\frac{1}{2(1+\nu)}\left( (1-\nu)\hat{c}_1 -\nu\left(\hat{c}_0+{\cal S}\right)+ {\cal R}  \right) \ .
\end{equation}
The unknowns $c$ and $d$ from the particular solution part can be chosen arbitrarily. Their influence
on the homogeneous solution part is restricted to $B_0$ and $D_0$ 
and is compensated in the complete solution.

\subsubsection{Summary: Coefficients of Homogeneous Solution}

For the sake of clearness the equations for the determination of the homogeneous solution part
as derived in Eqs.~(\ref{Eq:AjBj}), (\ref{Eq:CnDn}),
(\ref{Eq:B0}), (\ref{Eq:D0}), (\ref{Eq:DnBj_1}) and (\ref{Eq:DnBj_2}) are summarized:
\begin{eqnarray}
A_j \la_j & = & - B_j \left( 2\nu + \la_jL \coth(\la_j L)\right) \\
C_n k_n & = & -  D_n \left( (2-2\nu) + k_nR \frac{I_0(k_nR)}{I_1(k_nR)} \right) \\ 
D_n &=& - \left( k_nR \left(
\frac{I_0(k_nR)^2}{I_1(k_nR)^2}-1\right) - \frac{2-2\nu}{k_nR}\right)^{-1} \times \nonumber\\
& & \qquad \times \left(\frac{4 \hat{c}_2 (-1)^n}{k_n^2} 
+ \sum_{m=1}^{\infty} B_m (-1)^n \frac{4\la_m
k_n^2}{L(k_n^2+\la_m^2)^2} J_0(\la_m R)\right) \label{Eq:ggs_infinite}
 \\
B_j &=& \left(J_0(\la_jR) \left( \coth(\la_j L) + \frac{\la_j
L}{\sinh^2(\la_j L)} \right)\right)^{-1} \times \nonumber\\
& & \qquad \times 
\left(  \frac{2}{\la_j^2} \hat{c}_2 
-\sum_{k=1}^{\infty} (-1)^k D_k k_k \left( 
\frac{4\la_j^2}{R(k_k^2+\la_j^2)^2}  \right) \right) \label{Eq:ggs_infinite1}
 \\
B_0 &=& \frac{1}{6}\left(-\hat{c}_0-\hat{c}_1-{\cal S}\right) - D_0 \\
D_0 &=&\frac{1}{2(1+\nu)}\left( (1-\nu)\hat{c}_1 -\nu\left(\hat{c}_0+{\cal S}\right)+ {\cal R}\right) 
\end{eqnarray}
with
\begin{equation}
\hat{c}_0=\gamma \left(\frac{4c\nu}{1-\nu}+2(L^2+d)\right)
\, 
\hat{c}_1 = \gamma \left( \frac{(2\nu-3)}{4(1-\nu)}R^2 +\frac{2(c+\nu
d)}{1-\nu}\right) 
\,
\hat{c}_2 = \gamma \frac{2\nu }{1-\nu}
\end{equation}
and ${\cal S}={\cal R}+{\cal Z} $.

Now we have determined all unknown coefficients in Equation (\ref{Eq:chi}). Then,  following Eq.~(\ref{Eq:xiLove}) we can directly calculate the homogeneous displacement. By adding the homogeneous and the particular solution we get the total displacement of the cylinder.

\subsubsection{Convergence of the Homogeneous Solution}

The infinite system (\ref{Eq:ggs_infinite},\ref{Eq:ggs_infinite1}) can be approximately solved by reducing it to a finite system, that is, by expanding the sums only to $n=N$ and $j=J$. Then we have a system of $N+J$ equations. By increasing the values of $N$ and $J$ one can improve the accuracy of the solution and find their limits.

This approach is justified, since we can prove that these infinite sums do converge, that means, that the infinite system (\ref{Eq:ggs_infinite},\ref{Eq:ggs_infinite1}) possesses a unique bounded solution. In doing so we have to use the theory of regular infinite systems as has been done by \cite{Kovalenko} and \cite{Meleshko}.

In the following we will show first that the equation system (\ref{Eq:ggs_infinite},\ref{Eq:ggs_infinite1}) is a regular infinite system and second that it possesses a unique solution. For the sake of simplicity we introduce the abbreviations
\begin{eqnarray}
P_n & = & R^2 \left( \frac{I_0(k_nR)^2}{I_1(k_nR)^2}-1\right) - \frac{2-2\nu}{k_n^2} \\
Q_j & = & \frac{L}{\la_j} \left( \coth(\la_j L) + \frac{\la_j L}{\sinh^2(\la_j L)} \right) \\
X_n & = & D_n (-1)^n  \frac{k_n}{R} \\
Y_j & = & -B_j \frac{\la_j}{L} J_0(\la_j R) 
\end{eqnarray}
and rewrite the system (\ref{Eq:ggs_infinite},\ref{Eq:ggs_infinite1}) as
\begin{eqnarray}\label{Eq:ggs_meleshko}
X_n - \frac{1}{P_n} \sum_{j=1}^{\infty} \frac{4 k_n^2}{(k_n^2+\la_j^2)^2}  Y_j 
= X_n - \sum_{j=1}^{\infty} c_{n,j}  Y_j 
&=& -\frac{1}{P_n} \frac{4 \hat{c}_2}{k_n^2} \nonumber \\
Y_j - \frac{1}{Q_j} \sum_{n=1}^{\infty} \frac{4\la_j^2}{(k_n^2+\la_j^2)^2}  X_n 
= Y_j - \sum_{n=1}^{\infty} d_{j,n}  X_n 
&=& - \frac{1}{Q_j} \frac{2 \hat{c}_2 }{\la_j^2} \ .
\end{eqnarray}

The system (\ref{Eq:ggs_meleshko}) can be written in a combined form 
\begin{equation}
z_i =  \sum_{l=1}^{\infty} e_{i,l} z_l + b_i \, , \qquad i = 1, 2, \ldots, \infty \, ,\label{CombinedSystem}
\end{equation}
with $z_{2l-1}=X_l$ and $z_{2l}=Y_l$.
Thus the vector $z$ alternating contains the terms $X_n$ and $Y_j$ from Equation (\ref{Eq:ggs_meleshko}).
Then the matrix $e$ must fulfill
\begin{eqnarray}
e_{2m-1,2l-1} &=& 0\\
e_{2m-1,2l} &=& c_{m,l}\\
e_{2m,2l-1} &=& d_{m,l}\\
e_{2m,2l} &=& 0 
\end{eqnarray}
with
\begin{equation}
c_{n,j}:=\frac{1}{P_n} \frac{4 k_n^2}{(k_n^2+\la_j^2)^2} 
\qquad {\rm and} \qquad
d_{j,n}:=\frac{1}{Q_j} \frac{4\la_j^2}{(k_n^2+\la_j^2)^2} \ .
\end{equation}

An infinite system of the form (\ref{CombinedSystem}) is called {\it regular}, if in each equation of this system the sum of the norms of the coefficients is smaller than one (see \cite{Kantorowitsch})
\begin{equation}
\sum_{l=1}^{\infty} |e_{i,l}| < 1 \qquad (i=1,2,\dots) \ .
\end{equation}
Due to our substitutions we have 
\begin{eqnarray}
\sum_{l=1}^{\infty} |e_{i,l}|  &=& \sum_{l=1}^{\infty} |c_{m,l}| \qquad (i=2m-1) \\
\sum_{l=1}^{\infty} |e_{i,l}|  &=& \sum_{l=1}^{\infty} |d_{m,l}| \qquad (i=2m) 
\end{eqnarray}
so that the condition for regularity reads
\begin{eqnarray}
\sum_{l=1}^{\infty} |c_{m,l}| < 1 \\
\sum_{l=1}^{\infty} |d_{m,l}| < 1 \ .
\end{eqnarray}

These conditions are satisfied for the system (\ref{Eq:ggs_infinite},\ref{Eq:ggs_infinite1}) or (\ref{Eq:ggs_meleshko}) because we can calculate using (\ref{B:A9}) and (\ref{B:A7}) 
\begin{eqnarray}\label{Eq:PsiPhi}
\sum_{j=1}^{\infty} c_{n,j} = \frac{1}{P_n} \sum_{j=1}^{\infty} \frac{4 k_n^2}{(k_n^2+\la_j^2)^2} 
&=& \frac{1}{P_n} R^2 \left( \frac{I_0(k_nR)^2}{I_1(k_nR)^2}-1 \right) - \frac{4}{k_n^2} \nonumber \\
&=& \frac{1}{P_n} \left( P_n + \frac{2-2\nu}{k_n^2} \right) - \frac{4}{k_n^2} \nonumber\\
& = & 1 + \frac{1}{P_n} \left( \frac{2-2\nu-4}{k_n^2} \right) \nonumber \\
&=& 1-\Psi_n \qquad{\rm with}  \qquad \Psi_n = \frac{2+2\nu}{k_n^2 P_n} \nonumber  \\
\sum_{n=1}^{\infty} d_{j,n} = \frac{1}{Q_j} \sum_{n=1}^{\infty} \frac{4\la_j^2}{(k_n^2+\la_j^2)^2} 
&=&\frac{1}{Q_j} \frac{L}{\la_j} \left( \coth(\la_j L) + \frac{\la_j L}{\sinh^2(\la_j L)} - \frac{2}{\la_jL}\right)\nonumber \\
&=&\frac{1}{Q_j} Q_j - \frac{2}{Q_j \la_j^2} \nonumber\\
& = & 1 - \frac{2}{Q_j \la_j^2} \nonumber \\
&=& 1-\Phi_j \qquad {\rm with} \qquad \Phi_j = \frac{2}{\la_j^2 Q_j} \ .
\end{eqnarray}
Obviously, $Q_j>0$ as the cylinder length $L$ and the radius $R$ are positive and the zeros of the Bessel function of first order $\la_j R$ are positive, too. Using properties of the Bessel functions, it can be shown that also $P_n>0$. Therefore no singularities can occur. Furthermore $\Psi$ and $\Phi$ are positive definite series and none of their terms becomes one for all values
of $n$ and $j$, thus $1-\Psi_n$ and $1-\Phi_j$ are always smaller than zero. Therefore, (\ref{CombinedSystem}) is regular. 

The final conclusion that the regular infinite system (\ref{Eq:ggs_infinite},\ref{Eq:ggs_infinite1}) has a unique bounded solution can be inferred from a theorem \cite{Kantorowitsch}, that a regular infinite system (\ref{CombinedSystem}) whose free terms $b_i$ fulfill the condition
\begin{equation}
|b_i| \le K \rho_i
\qquad {\rm with} \qquad
\rho_i:= 1 - \sum_{l=1}^{\infty} |e_{i,l}|
\end{equation}
does have a bounded solution $|z_i| \le K$, where $K$ is an arbitrary constant.

From (\ref{Eq:PsiPhi}) its obvious that
$\rho_i = 1-|1-\Psi_i|=|\Psi_i|$ and $\rho_i = 1-|1-\Phi_i|=|\Phi_i|$.
Thus the condition
$|b_i| \le K \rho_i$ 
can be divided into two conditions
\begin{equation}
\left| -\frac{4 \hat{c}_2}{P_n k_n^2} \right| \le K |\Psi_n| 
= K_1 \left| \frac{2+2\nu}{k_n^2 P_n} \right|
\le K \left| \frac{2+2\nu}{k_n^2 P_n} \right|
\end{equation}
and
\begin{equation}
\left| -\frac{2 \hat{c}_2}{Q_j \la_j^2}\right| 
\le K |\Phi_j| = K_2 \left|\frac{2}{\la_j^2 Q_j} \right| 
\le K \left|\frac{2}{\la_j^2 Q_j} \right|\ .
\end{equation}
Since $K_1$ and $K_2$ are arbitrary constants they can be chosen to be
\begin{equation}
K_1 = \frac{4 \hat{c}_2}{2+2\nu}
\qquad {\rm and} \qquad
K_2 = \hat{c}_2
\end{equation}
which fulfills the above condition. One can choose $K$ to be $K=K_1=\frac{4 \hat{c}_2}{2+2\nu}$, 
as the Poisson number $\nu$ lies between 0 and 0.5 and therefore $K_1 > K_2$. Thus we have proven that the condition for the existence of a bounded solution is fulfilled.


\section{Comparison between Analytical and FEM Solution} 

Having proven that the infinite equation system (\ref{Eq:ggs_infinite},\ref{Eq:ggs_infinite1}) possesses a unique bounded
solution, we can now numerically calculate this solution with arbitrary order of accuracy by expanding the infinite series to sufficiently large numbers $N=J$. Although the infinite series converge very quickly, an expansion 
to higher orders $N=J$ still gives an improvement of accuracy which can be seen 
in Table~\ref{Tab:vgl_NJ}.

\begin{table}[b]
\caption{\label{Tab:vgl_NJ}Comparison of the analytical solution at point $r=1=R$, $z=2=L$ for different expansion orders $N=J$ of the infinite sums in equation system (\ref{Eq:ggs_infinite}).}
\begin{center}
\begin{tabular}{@{}ccc}
\br
$N=J$    & $\xi_r$   & $\xi_z$ \\
\mr
100 &  $-6.471\cdot10^{-15}$ & $1.637\cdot10^{-13}$\\
1000 & $-6.233\cdot10^{-15}$ & $1.629\cdot10^{-13}$\\
1700 & $-6.218\cdot10^{-15}$ & $1.628\cdot10^{-13}$\\
\br
\end{tabular}
\end{center}
\end{table}

\begin{figure}[h]
\centering\epsfig{file=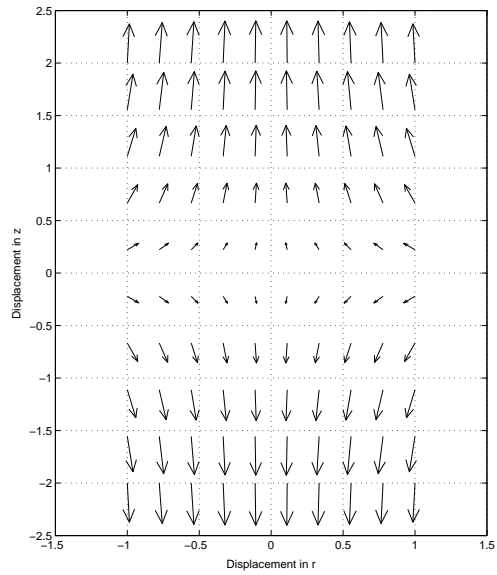,width=0.5\textwidth}
\caption{\label{Fig:gg_displfield}Analytical solution: Displacement field of a cylinder under influence of 
a spherical tidal gravitational force field. The displacements are plotted over the body coordinates $r$ and $z$. The cylinder boundaries are at $z=\pm L=\pm 2$ and $r=R=1$.}
\end{figure}

Fig.~\ref{Fig:gg_displfield} shows the resulting total displacement field from the analytical solution of Equation (\ref{Eq:xi_total}). The infinite sums in the analytical solution were expanded to $N=J=1700$.

\begin{figure}[h]
 \centering\epsfig{file=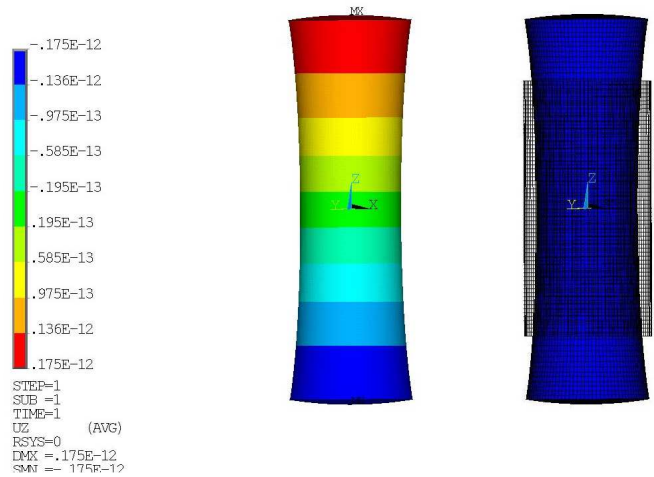,angle=-90,width=0.7\textwidth}
\caption{\label{Fig:ansys_zyl}Finite element solution: Deformation of the cylinder 
under influence of a spherical tidal gravitational force field. The deformation is scaled by a factor 
of $6\cdot10^{13}$. Right: Deformed cylinder shape and original finite element mesh. 
Left: the scale shows the $z$ displacements.}
\end{figure}

Now we can compare our analytical solution for our simplified problem with the result obtained with help of a finite element code in order to verify its applicability. 
The finite element analysis was done with the commercial FEM code ANSYS. The cylinder model of radius $R=1$ and length $L=2R$ was divided into approximately 110\,000 hexahedron elements. Hexahedron elements allow the creation of a structured finite element mesh which ensures a high relative accuracy of the finite element solution. Fig.~\ref{Fig:ansys_zyl} shows the deformation of the cylinder 
in the tidal gravitational force field as result of the FEM calculation.

Table \ref{Tab:vgl} contains a quantitative comparison between the displacements from the analytical and FEM solution for special points of the cylinder. Analytical as well as FEM solution are completely symmetric, i.e. the norms of the displacements of two opposing points of the cylinder are equal. Please note that we have chosen a very fine mesh for the finite element solution in order to get high accuracy.

\begin{table}
\caption{\label{Tab:vgl}Displacements $\xi_r$ and $\xi_z$ in $r$ and $z$ direction; Comparison between analytical and FEM solution. The cylinder boundaries are $r=R=1$ and $z=\pm L=\pm 2$. In the analytical solution the infinite series were expanded to $N=J=1700$. 
The constant $\gamma:=-\frac{GM_\oplus}{2 r_{M}^3}\rho=-\frac{3.986004415\cdot10^{14}}{2\cdot343\cdot10^{18}}2350$ which corresponds to a 7\,000 km Earth orbit.}
\begin{indented}
\item[]\begin{tabular}{@{}cccccc}
\br
\multicolumn{2}{c}{}    & \multicolumn{2}{c}{Analytical} & \multicolumn{2}{c}{Numerical}\\
\hline
$r$    &   $z$    & $\xi_r$   & $\xi_z$ &  $\xi_r$   & $\xi_z$ \\
\hline
1               &2& $-6.218\cdot10^{-15}$& $1.628\cdot10^{-13}$&$-6.202\cdot10^{-15}$&$1.625\cdot10^{-13}$\\
1               &0& $-3.491\cdot10^{-14}$& 0                   &$-3.486\cdot10^{-14}$&$-1.66\cdot10^{-23}$\\ 
$0.49507\sqrt{2}$&1&$-2.059\cdot10^{-14}$& $1.151\cdot10^{-13}$&$-2.062\cdot10^{-14}$&$1.150\cdot10^{-13}$\\
$0.11888\sqrt{2}$&1&$-5.208\cdot10^{-15}$& $1.183\cdot10^{-13}$&$-5.236\cdot10^{-15}$&$1.182\cdot10^{-13}$\\
\br
\end{tabular}
\end{indented}
\end{table}

Obviously, the FEM solution and the analytical solution agree very well. The small differences are based on the nature of the FEM analysis. The elements in which the cylinder is divided cannot be chosen to be infinite small, they are 'finite' and thus the FEM solution is a kind of summation over all elements of finite size instead of an integration where the limit to infinitesimal small element size can be performed. Furthermore during the FEM analysis at least three points must be fixed in order to prevent the cylinder from rigid body rotations. 
In the current FEM analysis the center of mass as well as four of the next nodes were fixed in order to assure the perfect symmetry of the FEM solution.


\section{Summary and Outlook}

An analytical solution for the problem of an elastic, isotropic, homogeneous freely--flying cylinder in space under the influence of 
a tidal gravitational force has been worked out. The motivation for this work was the verification of the use of FEM codes for modeling the deformations of optical resonators in high precision space experiments. It could be shown that the analytical and the FEM solutions are equal within the limits set by the nature of the different solution approaches. The analytical solution itself gives a new application field for the theory of elasticity as so far (according to the knowledge of the authors) 
no examples with tidal gravitational forces have been calculated. 

A further application case of such analytical solutions is the estimation of the noise
due to thermoelastic deformations in gravitational wave detectors. In particular with respect to the upcoming 
LISA (Laser Interferometer Space Antenna) mission \cite{danzmann03:lisa} these investigations are of great interest. Some groups already dealt with this problem, see \cite{BHV,LiuThorne}. However, the solutions are given for infinite half--spaces only or some approximations have to be included in order to satisfy all boundary conditions. Therefore we are currently working on an exact analytical solution for the estimation of thermoelastic noise in gravitational wave detectors.

\section{Acknowledgments}

We like to thank H. Dittus and H. Kienzler and his team for fruitful discussions. Special thanks to Reinhard Ristau for the profound introduction into ANSYS programming, and to Eva Hackmann for her extensive literature search and mathematical advice. Financial support of the German Aerospace Agency DLR is gratefully acknowledged. 

\appendix

\section{Bessel function relations}\label{A:bessel}

For convenience of the reader we collect some equations from \cite{GradshteynRyzhik}, \cite{AbramowitzStegun}, \cite{Arfken}, \cite{Watson} used within this article.

The Bessel functions $J_{\nu}(x)$ of first kind and $\nu$th order
\begin{equation}\label{Eq:Bessel}
J_{\nu}(x) = \sum_{v=0}^{\infty} \frac{(-1)^v}{v! \Gamma(\nu+v+1)} \left( \frac{x}{2}\right)^{\nu+2v} 
\qquad \nu \in \mathbb{R}
\end{equation}
are solutions of the Bessel differential equation
\begin{equation}
x^2 y'' + x y' + (x^2-\nu^2) y = 0 \ .
\end{equation}
Functions with the argument $l x$ fulfill the differential equation
\begin{equation}\label{Eq:A_bessel_l}
x^2 y'' + x y' + (l^2 x^2-n^2) y = 0 \ .
\end{equation}
The modified Bessel functions $I_{\nu}(x)$ of first kind and $\nu$th order are defined as
\begin{equation}\label{Eq:Bessel_I} 
I_{\nu}(x) = i^{-{\nu}} J_{\nu}(ix) = \sum_{v=0}^{\infty} \frac{1}{v! \Gamma(\nu+v+1)} \left( \frac{x}{2}\right)^{\nu+2v} \qquad \nu \in  \mathbb{R} \ . 
\end{equation}
The orthogonality relations for the Bessel functions can be derived as
\begin{equation}\label{Eq:A_arfken1150}
\frac{R^2}{2} \left( J_{\nu+1}(\zeta_{\nu m}) \right)^2 \delta_{m n} 
= \int_0^R J_{\nu}(l_{\nu m} r) J_{\nu}(l_{\nu n} r) r dr
\end{equation}
for $\nu>-1, \, \nu \in \mathbb{R}$ provided that $\zeta_{\nu m} = l_{\nu m} R$ 
and $\zeta_{\nu n} = l_{\nu n} R$ are the $m$th respectively the $n$th zero of $J_{\nu}$, i.e. 
$J_{\nu}(l_{\nu m} R)=0$ and $J_{\nu}(l_{\nu n} R)=0$ 
and it is shown that they form a complete set, so every arbitrary
function $f(r)$ can be represented by a Bessel-Fourier series
\begin{equation}\label{Eq:A_arfken1151}
f(r) = \sum_{m=1}^{\infty} c_{\nu m} J_{\nu} \left( l_{\nu m} \frac{r}{R}\right)
\end{equation}
for $0 \le r \le R, \nu > -1$.
The coefficients can be determined via
\begin{equation}\label{Eq:A_arfken1152}
c_{\nu m} = \frac{2}{R^2 [J_{\nu+1}(\zeta_{\nu m})]^2} 
\int_0^R f(r) J_{\nu}\left( l_{\nu m} \frac{r}{R}\right) r dr \ .
\end{equation}

Under the condition that $l_{\nu m}$ is related to the $m$th zero $\zeta_{\nu m}$ of $\frac{\partial}{\partial r}
J_{\nu}(l_{\nu m} r)$ via $\zeta_{\nu m}=l_{\nu m}R$, i.e. $\frac{\partial}{\partial r}
J_{\nu}(l_{\nu m} r)|_{r=R} =0$, a second orthogonality relation for Bessel functions is
\begin{equation}\label{Eq:A_arfken1123}
\frac{R^2}{2} \left(1-\frac{\nu^2}{l_{\nu m}^2}\right) \left( J_{\nu}(\zeta_{\nu m}) \right)^2 \delta_{m n} 
= \int_0^R J_{\nu}(l_{\nu m} r) J_{\nu}(l_{\nu n} r) r dr
\end{equation}
for $\nu>-1\nu>-1, \, \nu \in \mathbb{R}$.

Because of this second orthogonality relation one can represent an arbitrary function as expansion of a
so-called Dini series
\begin{equation}
f(r) = \sum_{m=1}^{\infty} d_{\nu m} J_{\nu} \left( l_{\nu m} \frac{r}{R}\right)
\end{equation}
\begin{equation}\label{Eq:A_arfken1126b}
d_{\nu m} = \frac{2}{R^2 (1-\nu^2/\zeta_m^2)[J_{\nu}(\zeta_{\nu m})]^2} 
\int_0^R f(r) J_{\nu}\left( l_{\nu m} \frac{r}{R}\right) r dr \ .
\end{equation}

Note that in the case of a Dini series for $\nu=0$ an additional term $d_0$ has to be added in the series expansion \cite{Watson}.
For Dini series expansions and $\nu=0$ one has
\begin{equation}\label{Eq:A_dini}
f(r) = d_0 + \sum_{m=1}^{\infty} d_{m} J_{0} \left( l_{m} \frac{r}{R}\right)
\end{equation}
with
\begin{equation}
d_{m} = \frac{2}{R^2 [J_{0}(\zeta_{m})]^2} 
\int_0^R f(r) J_{0}\left( l_{m} \frac{r}{R}\right) r dr 
\end{equation}
and
\begin{equation}
d_0 = \frac{2}{R^2} \int_0^R f(r) r dr 
\end{equation}
where $\zeta_m=l_mR$ is the $m$th zero of $J_1$.

\smallskip
Provided that $l_m = \zeta_m/R$, where $\zeta_m$ is the $m$th zero of 
$J_1$ and $R$ is the maximum value of $r$ one can derive the special relations
\begin{equation}\label{Eq:A_thorne41}
\frac{R^2}{2} \left(J_0(\zeta_{m})\right)^2 \delta_{m n} 
= \int_0^R J_1(l_m r) J_1(l_n r) r dr = \frac{R^2}{2} \left(J_2(\zeta_{m})\right)^2 \delta_{m n} \ ,
\end{equation}
\begin{equation}\label{Eq:A_thorne42}
\frac{R^2}{2} \left(J_0(\zeta_{m})\right)^2 \delta_{m n} 
= \int_0^R J_0(l_m r) J_0(l_n r) r dr \ ,
\end{equation}
\begin{equation}\label{Eq:A_thorne43}
0 = \int_0^R J_0(l_m r)  r dr \ .
\end{equation}

\section{Relations of Fourier and Dini Series}\label{B:fourier}

The orthogonality relations for Fourier series are
\begin{eqnarray}
\int_{-\pi}^{\pi} \sin(n'z) \sin(nz) dz &=&
\left\{ \begin{array}{ll}
	 \pi \delta_{nn'}  & n \neq 0\\
	 0                & n=0
        \end{array} \right. \nonumber \\
\int_{-\pi}^{\pi} \cos(n'z) \cos(nz) dz &=&
\left\{ \begin{array}{ll}
	 \pi \delta_{nn'}  & n \neq 0\\
	 2\pi                & n=0
        \end{array} \right. \nonumber \\
\int_{-\pi}^{\pi} \sin(n'z) \cos(nz) dz &=& 0 
\end{eqnarray}
for integer $n$ and $n'$, $n'\neq0$.

We also have 
\begin{eqnarray}\label{B:ortho}
\int_{-L}^{L} \sin(k_{n'z}) \sin(k_nz) dz &=&
\left\{ \begin{array}{ll}
	 L \delta_{nn'}  & n \neq 0\\
	 0                & n=0
        \end{array} \right. \nonumber \\
\int_{-L}^{L} \cos(k_{n'}z) \cos(k_nz) dz &=&
\left\{ \begin{array}{ll}
	 L \delta_{nn'}  & n \neq 0\\
	 2L                & n=0
        \end{array} \right. \nonumber \\
\int_{-L}^{L} \sin(k_{n'}z) \cos(k_nz) dz &=& 0 
\end{eqnarray}
for $k_n=n\pi/L$ and $k_{n'}=n'\pi/L$, $n, n'$ are integer numbers, $n'\neq0$.

As the trigonometric functions form a complete orthogonal set, each function $f(z)$ 
can be represented by a Fourier series
\begin{equation}\label{B:Fseries}
f(z) = \frac{a_0}{2}+\sum_{n=1}^{\infty} a_n \cos(k_n z) + \sum_{n=1}^{\infty} b_n \sin(k_n z) \ .
\end{equation}
The Fourier coefficients are determined via
\begin{equation}\label{B:Fcoeff}
a_n =\frac{1}{L} \int_{-L}^{L} f(z) \cos\left(\frac{n\pi}{L}z\right)dz \, , \qquad  
b_n = \frac{1}{L} \int_{-L}^{L} f(z) \sin\left(\frac{n\pi}{L}z\right)dz \ .
\end{equation}

\medskip
In the following we collect some Fourier and Dini series expansions used within the article \cite{GradshteynRyzhik}, \cite{Meleshko}, for $k_n=n\pi/L$ and $J_1(\la_jR)=0$, integer $n,j$:
\begin{equation}\label{B:A1}
\la z \frac{\sinh(\la z)}{\sinh(\la L)}+(1-\la L \coth(\la L)) \frac{\cosh(\la z)}{\sinh(\la L)}
= \sum_{n=1}^{\infty} (-1)^n \frac{4\la k_n^2}{L(k_n^2+\la^2)^2} \cos(k_n z)
\end{equation}
\begin{equation}\label{B:A4}
r \frac{I_1(kr)}{I_1(kR)} + \left( \frac{2}{k} - R \frac{I_0(kR)}{I_1(kR)}\right) \frac{I_0(kr)}{I_1(kR)}
= \sum_{j=1}^{\infty} \frac{4\la_j^2}{R(k^2+\la_j^2)^2} \frac{J_0(\la_j r)}{J_0(\la_j R)}
\end{equation}
\begin{equation}\label{B:A5}
\frac{I_0(kr)}{I_1(kR)} 
= \frac{2}{kR} + \sum_{j=1}^{\infty} \frac{2k}{R(k^2+\la_j^2)} \frac{J_0(\la_j r)}{J_0(\la_j R)}
\end{equation}
From these equations one can derive the following sums, by setting $z=L$ and $r=R$:
\begin{equation}\label{B:A7}
\sum_{n=1}^{\infty} \frac{4 \la^2}{(k_n^2+\la^2)^2} = 
\frac{L}{\la} \left( \coth(\la L) + \frac{\la L}{\sinh^2(\la L)} - \frac{2}{\la L}\right) 
\end{equation}

\begin{equation}\label{B:A10}
\sum_{j=1}^{\infty} \frac{4 \la_j^2}{(k^2+\la_j^2)^2}
= R^2 \left(1 - \frac{I_0(kR)^2}{I_1(kR)^2}\right) + \frac{2RI_0(kR)}{kI_1(kR)}
\end{equation}

\begin{equation}\label{B:A9}
\sum_{j=1}^{\infty} \frac{4 k^2}{(k^2+\la_j^2)^2}
= R^2 \left( \frac{I_0(kR)^2}{I_1(kR)^2}-1 \right) - \frac{4}{k^2}
\end{equation}
If $k \to 0$ in Equation (\ref{B:A10}) one gets
\begin{equation}\label{B:A11}
\sum_{j=1}^{\infty} \frac{1}{\la_j^2} = \frac{R^2}{8} \ . 
\end{equation}
In particular, 
\begin{equation}\label{B:A11_var}
\sum_{j=1}^{\infty} \frac{1}{J_0(\la_jR) \la_j^2} = -\frac{R^2}{8} \ . 
\end{equation}


\section*{References}
\bibliographystyle{plain}
\bibliography{./references}

\begin{thebibliography}{10}

\bibitem{AbramowitzStegun}
M.~Abramowitz and I.~Stegun.
\newblock {\em {Handbook of Mathematical Functions}}.
\newblock Dover Publications, Inc, New York, 1972.

\bibitem{Arfken}
G.~Arfken.
\newblock {\em {Mathematical Methods for Physicists}}.
\newblock Academic Press, Inc, San Diego, 1985.

\bibitem{BHV}
F.~Bondu, P.~Hello, and J.-Y. Vinet.
\newblock {Thermal noise in mirrors of interferometric gravitational wave
  antennas}.
\newblock {\em Physical Letters A}, 246:227, 1998.

\bibitem{Boresi}
A.P. Boresi.
\newblock {\em {Elasticity in Engineering Mechanics}}.
\newblock Elsevier Science Publishing Co, 1987.

\bibitem{Braxmaieretal02}
C.~Braxmaier, H.~M{\"u}ller, O.~Pradl, J.~Mlynek, A.~Peters, and S.~Schiller.
\newblock Test of relativity using a cryogenic optical resolator.
\newblock {\em Phys. Rev. Lett.}, 88:010401, 2002.

\bibitem{danzmann03:lisa}
K.~Danzmann and A.~R\"udiger.
\newblock {LISA technology -- concept, status, prospects}.
\newblock {\em Classical and Quantum Gravity}, 20(10):S1--S9, 2003.

\bibitem{Muelleretal03}
H.~M{\"u}ller et~al.
\newblock {Modern {M}ichelson--{M}orley experiment using cryogenic optical
  resonators}.
\newblock {\em Phys. Rev. Lett.}, 91(020401), 2003.

\bibitem{Wolfetal03}
P.~Wolf et~al.
\newblock {Tests of relativity using a microwave resonator}.
\newblock {\em Phys. Rev. Lett.}, 90:060402, 2003.

\bibitem{GradshteynRyzhik}
I.S.. Gradshteyn and I.M. Ryzhik.
\newblock {\em {Tables of Integrals, Series and Products}}.
\newblock Academic Press, Inc, San Diego, 1980.

\bibitem{Kantorowitsch}
L.W. Kantorowitsch and W.I. Krylow.
\newblock {\em {N\"aherungsmethoden der h\"oheren Analysis}}.
\newblock VEB Deutscher Verlag der Wissenschaften, Berlin, 1956.

\bibitem{Kienzler}
R.~Kienzler.
\newblock {Eine vollst\"andige Gleichungsstruktur der linearen
  Elastizit\"atstheorie}.
\newblock {\em Ingenieur-Archiv}, 51:421--426, 1982.

\bibitem{Kovalenko}
A.~D. Kovalenko.
\newblock {\em {Thermoelasticity -- Basic Theory and Applications}}.
\newblock Wolters-Noordhoff Publishing Groningen, 1969.

\bibitem{Laemmerzahletal04}
C.~L\"ammerzahl, G.~Ahlers, N.~Ashby, M.~Barmatz, P.L. Biermann, H.~Dittus,
  V.~Dohm, R.~Duncan, K.~Gibble, J.~Lipa, N.A. Lockerbie, N.~Mulders, and
  C.~Salomon.
\newblock {Experiments in {F}undamental {P}hysics scheduled and in development
  for the {ISS}}.
\newblock {\em Gen. Rel. Grav.}, 36:615, 2004.

\bibitem{LandauLifschitz}
L.D. Landau and E.M. Lifschitz.
\newblock {\em {Lehrbuch der Theoretischen Physik - Bd 7 Elastizitätstheorie}}.
\newblock Akademie-Verlag Berlin, 1966.

\bibitem{Leipholz}
H.~Leipholz.
\newblock {\em {Theory of Elasticity}}.
\newblock Noordhoff International Publishing, Leyden, 1974.

\bibitem{LiuThorne}
Y.T. Liu and K.~Thorne.
\newblock {Thermoelastic noise and homogeneous thermal noise in finite sized
  gravitational-wave test masses}.
\newblock {\em Physical Review D}, 62:122002, 2000.

\bibitem{Love}
A.E.H. Love.
\newblock {\em {A Treatise on the Mathematical Theory of Elasticity}}.
\newblock 4th edn. Cambridge University Press, Cambridge, 1927.

\bibitem{Lurje}
A.I. Lurje.
\newblock {\em {Räumliche Probleme der Elastizitätstheorie}}.
\newblock Akademie-Verlag, Berlin, 1963.

\bibitem{laemmerzahl04:_optis}
C.~Lämmerzahl, I.~Ciufolini, H.~Dittus, L.~Iorio, H.~Müller, A.~Peters,
  E.~Samain, S.~Scheithauer, and S.~Schiller.
\newblock {OPTIS -- An Einstein Mission for Improved Tests of Special and
  General Relativity}.
\newblock {\em General Relativity and Gravitation}, 36(10), 2004.

\bibitem{Marsden}
J.E. Marsden and T.J.R. Hughes.
\newblock {\em {Mathematical Foundations of Elasticity}}.
\newblock Dover Publications, Inc, New York, 1983.

\bibitem{Meleshko}
V.V. Meleshko.
\newblock {Equilibrium of an elastic finite cylinder: Filon's problem
  revisted}.
\newblock {\em Journal of Engineering Mathematics}, 16:355--376, 2003.

\bibitem{Watson}
G.N. Watson.
\newblock {\em {Theory of Bessel Functions}}.
\newblock Cambridge University Press, Cambridge, 1962.

\end{thebibliography}

\end{document}